\documentclass[preprint,review,12pt]{elsarticle}

\usepackage{graphicx}
\usepackage{amssymb}
\usepackage{newtxtext,newtxmath,amsmath,natbib}
\usepackage{xcolor}
\usepackage{lineno}

\journal{Icarus}

\begin{document}

\begin{frontmatter}


\title{Mars' atmospheric neon suggests volatile-rich primitive mantle}



\author[label1]{Hiroyuki Kurokawa}
\author[label2]{Yayoi N. Miura}
\author[label3]{Seiji Sugita}
\author[label3]{Yuichiro Cho}
\author[label4]{Fran\c{c}ois Leblanc}
\author[label5]{Naoki Terada}
\author[label5]{Hiromu Nakagawa}

\address[label1]{Earth-Life Science Institute, Tokyo Institute of Technology}
\address[label2]{Earthquake Research Institute, The University of Tokyo}
\address[label3]{The University of Tokyo}
\address[label4]{LATMOS/IPSL, Sorbonne Universit\'{e}, UVSQ, CNRS}
\address[label5]{Tohoku University}

% WORD LIMIT
% total 6500 words in the main text (introduction to conclusions)
% total 10 figures and tables
% total 50 references

\begin{abstract}

Martian atmospheric neon (Ne) has been detected by Viking and also found as trapped gas in Martian meteorites, though its abundance and isotopic composition have not been well determined. Because the timescale of Ne loss via atmospheric escape estimated from recent measurements with MAVEN is short (0.6--1 $\times$ 10$^8$ years), the abundance and isotope composition of Martian atmospheric Ne reflect recent atmospheric gas supply mostly from volcanic degassing. Thus, it can serve as a probe for the volatile content of the interior. Here we show that the tentatively-informed atmospheric Ne abundance suggests recent active volcanism and the mantle being richer in Ne than Earth's mantle today by more than a factor of 5--80. The estimated mantle Ne abundance requires efficient solar nebular gas capture or accretion of Ne-rich materials such as solar-wind-implanted dust in the planet formation stage, both of which provide important constraints on the abundance of other volatile elements in the interior and the accretion history of Mars. More precise determination of atmospheric Ne abundance and isotopic composition by in situ analysis or Mars sample return is crucial for distinguishing the possible origins of Ne.

\end{abstract}


\end{frontmatter}
%\begin{linenumbers}
%\linenumbers

\section{Introduction}
\label{sec:introduction}

Planetary interiors record their formation history and influence surface environments through volcanic and tectonic activities. The potentially large volatile content of the mantle is thought to have controlled the climate evolution of early Mars through volcanic supply of atmospheric gases and water \citep[e.g.,][]{Craddock+Greeley2009,Halevy+2007,Tian+2010,Sholes+2017,Ramirez+2014,Wordsworth+2017}. Thanks to exploration missions, the surface environment of Mars and its evolution has been enormously studied. In contrast, its interior is less understood. Given the current surface environment being highly modified by exogenous (impacts and atmospheric escape) and endogenous (volcanic degassing) processes through 4.5 billion years evolution \citep[e.g.,][]{Melosh+Vickery1989,Jakosky+1994,Pepin1991,Pepin1994,Terada+2009,Kurokawa+2014,Kurokawa+2016,Kurokawa+2018,Sakuraba+2019}, understanding its interior, which can still possess primitive information, is crucial to unveil the accretion and differentiation history. 

Noble gases are chemically inert, and thus useful to study both the origins of planetary volatile elements and atmospheric evolution \citep[e.g.,][]{Lammer+2020}. The Viking landers detected Martian atmospheric neon (Ne), argon (Ar), krypton (Kr), and xenon (Xe) \citep{Owen+1977}. The measured noble gas abundance was consistent with trapped gas in the Antarctic meteorite Elephant Moraine (EETA) 79001, which confirmed Martian origin of the meteorite \citep{Pepin1985}. Isotopic compositions of Ar, Kr, and Xe in the Martian atmosphere have been measured in situ by Viking and Curiosity \citep{Atreya+2013,Conrad+2016} and recovered from Martian meteorites \citep[][and references therein]{Smith+2020}. Those heavy noble gas isotopes have been used to constrain atmospheric evolution in the distant past \citep[>1 Ga,][]{Pepin1991,Pepin1994,Slipski+Jakosky2016,Jakosky+2017,Kurokawa+2018}. Helium (He) has not been found both in situ and from trapped gas in Martian meteorites, likely due to its low abundance \citep{Krasnopolsky+Gladstone2005,Smith+2020}. Ne isotope ratio ($^{20}$Ne/$^{22}$Ne) has been recovered from trapped gas in Martian meteorites, yet it has not reached consensus \citep{Swindle+1986,Wiens+1986,Garrison+1998,Park+Nagao2006,Park+2017,Smith+2020}. Although no in situ measurement is available so far, the determination of Ne isotopes by a future mission has been proposed \citep{Miura+2020}.

Here we show that Martian atmospheric Ne is a powerful probe for the volatile content of its interior and the planet formation history. Section 2 presents the balance between loss and supply of atmospheric Ne and shows the estimate of volcanic degassing flux and mantle Ne abundance. Section 3 presents models for Ne trapping into the mantle in the magma ocean stage to estimate the amount of Ne required in Mars formation stage. Section 4 presents the possible origins of Martian Ne: solar nebula gas, solar-wind-implanted dust, and chondritic materials. Section 5 discuss the requirements for precise determination of Ne abundance and isotopic composition by future missions, constraints from other volatile elements, and the comparison to Earth. We conclude in Section 6.

%He abundance constrained from EUV observations is explained by the balance between the loss and radioactive production and solar wind capture \citep{Krasnopolsky+Gladstone2005}. Degassing contribution is believed to be small from He \citep{Krasnopolsky+Gladstone2005}. 

%\note{\citet{Miura+2020} permeable membrane to measure Ne isotopes. }

\section{Loss and supply of atmospheric Ne on current Mars}
\label{sec:atmosphere}

\begin{figure}
    \centering
    \includegraphics[width=1.0\linewidth]{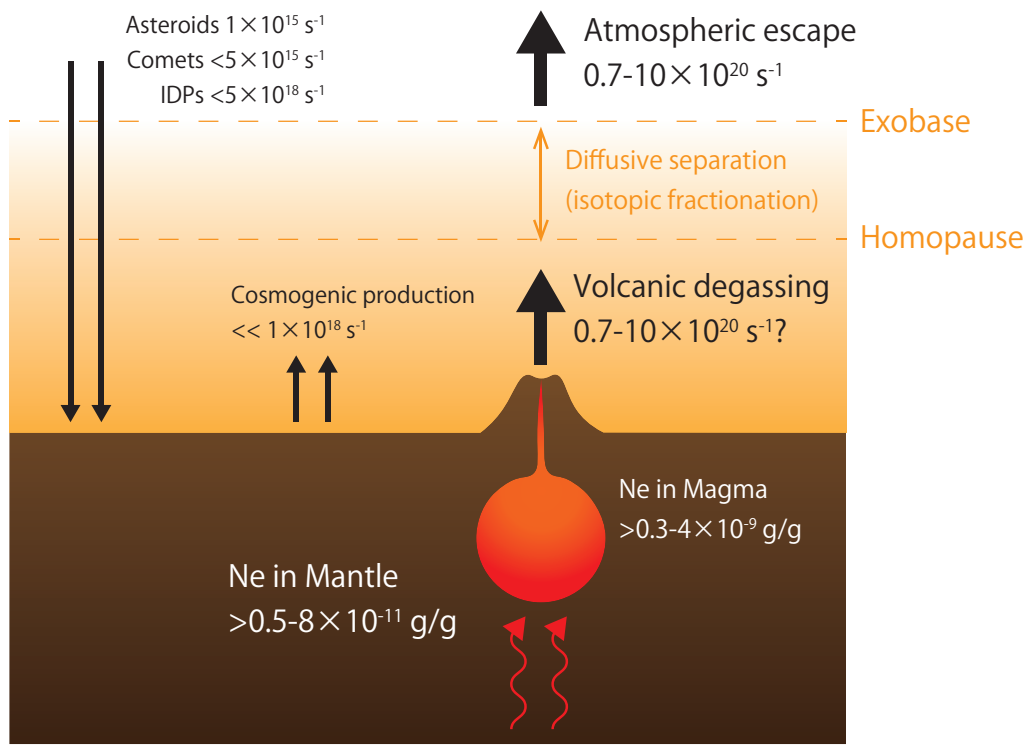}
    \caption{A schematic overview of loss and supply of Ne in the Martian atmosphere. See text for details.}
    \label{fig:overview}
\end{figure}

\begin{figure}
    \centering
    \includegraphics[width=1.0\linewidth]{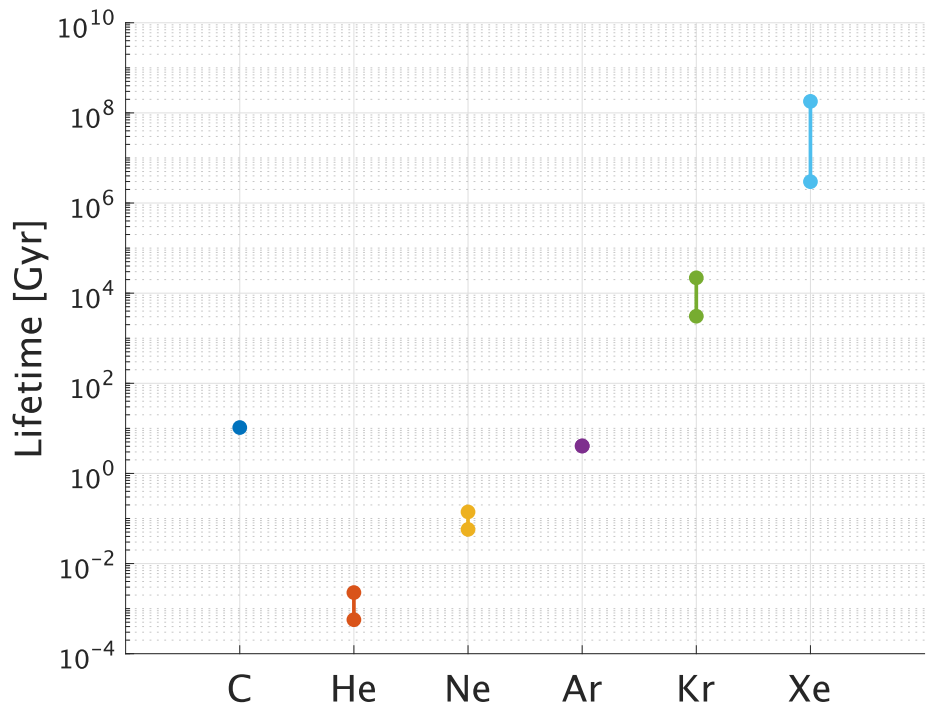}
    \caption{Lifetime of atmospheric gases against atmospheric escape on present-day Mars. The lifetime of C is estimated with the loss rates via sputtering (see text) and photochemical escape \citep{Hu+2015}. The lifetime of He is computed with the estimates of the total loss rate and abundance \citep{Krasnopolsky+Gladstone2005}. The lifetimes of noble gases other than He are computed from that of Ar with Equation \ref{eq:lifetime}.}
    \label{fig:lifetime}
\end{figure}

\subsection{Current knowledge on Ne abundance and isotope ratio}
\label{subsec:knowledge}

In situ measurement data of Ne abundance were only reported from Viking mission. Viking mass spectrometers detected $m/e=22$ (where $m$ and $e$ are the mass and charge numbers) signal and extraction of $^{44}$CO$_2^{++}$ component derived Martian atmospheric $^{22}$Ne abundance to be 0.25 ppm \citep{Owen+1977}. Assuming $^{20}$Ne/$^{22}$Ne = 10$\pm$3 and including the other uncertainties, $^{20}$Ne abundance was estimated to be 2.5$^{+3.5}_{-1.5}$ ppm \citep{Owen+1977}. The abundances of Ne relative to other noble gases recovered from Viking \citep{Owen+1977} and Martian meteorite \citep{Pepin1985} are consistent with each other, supporting the validity of the estimates.

Ne isotope ratio of the Martian atmosphere has been reported as an endmember of trapped gas in Martian meteorites. $^{20}$Ne/$^{22}$Ne ratios of 10.1$\pm$0.7 \citep{Wiens+1986} and 10.6$\pm$0.6 \citep{Swindle+1986} have been reported from the analysis of EETA 79001. In contrast, lower values, $^{20}$Ne/$^{22}$Ne = 7 \citep{Garrison+1998} and 7.3$\pm$0.2 \citep{Park+Nagao2006,Park+2017} have been also reported from the Yamato (Y) 793605 meteorite and from Dhofar 378 and Northwest Africa 7034, respectively. 

While a non-cosmogenic endmember of Ne in Martian meteorites is widely thought to originate from the Martian atmosphere \citep[e.g.,][]{Smith+2020}, Martian meteorites potentially contain Ne from the Martian interior as another endmember \citep{Mohapatra+2009}, as proposed for Ar and Xe \citep{Ott1988,Mathew+Marti2001,Swindle2002,Schwenzer+2007,Filiberto+2016}. However, both the isotopic ratio and abundance of potentially mantle-originated Ne are not well established. The dominance of cosmogenic Ne in Martian meteorite samples makes it difficult to understand the origins of trapped Ne components \citep[e.g.,][]{Smith+2020}. 

Though more precise determination is desired, some important conclusions can be derived from the reported values of Martian atmospheric Ne abundance and $^{20}$Ne/$^{22}$Ne ratio as below. 
%Dhofar (Dho) 378 and Northwest Africa (NWA)

\subsection{Atmospheric escape}
\label{subsec:escape}

The Martian atmosphere is being continuously lost by atmospheric escape processes \citep[e.g.,][]{Jakosky+2018}. Noble gases including Ne are removed chiefly by a process called sputtering (Figure \ref{fig:overview}): ejection of neutral atoms and molecules from near the exobase via collisions with pickup ions accelerated by the solar wind magnetic field \citep{Luhmann+Kozyra1991,Luhmann+1992}. As both Ne and Ar are removed by sputtering, we estimate the loss rate of Ne by relating the loss rate with that of Ar (4.9 $\times$ 10$^{22}$ s$^{-1}$) derived with a recent 3D simulation \citep{Leblanc+2018} which was validated by exospheric observations with The Mars Atmosphere and Volatile Evolution (MAVEN) spacecraft \citep{Leblanc+2019}. The relation of sputtering loss rates of two species $i, j$ denoted by $F_{i, {\rm sp}}$, $F_{j, {\rm sp}}$ can be well approximated by \citep{Johnson+2000,Leblanc+2012},
\begin{equation}
    \frac{F_{i, {\rm sp}}}{F_{\rm j, sp}} = \frac{Y_i}{Y_j} = \biggl( \frac{x_i}{x_j} \biggr)_{\rm exo} \frac{U_{{\rm es},j}}{U_{{\rm es},i}} = R_{{\rm diff}, i/j} \biggl( \frac{x_i}{x_j} \biggr)_{\rm homo} \frac{U_{{\rm es},j}}{U_{{\rm es},i}}, \label{eq:Fsp}
\end{equation}
where $Y_i$ is the sputtering yield (the number of particles removed per incident ion), $x_i$ is the molar fraction where the subscripts homo and exo denote the homopause and exobase, $U_{{\rm es},i}$ is the escape energy, and $R_{{\rm diff}, i/j}$ is the fractionation factor by diffusive separation between the homopause and exobase (i.e., the difference in the scale heights above the homopause), respectively. The diffusive separation factor is given by,
\begin{equation}
    R_{{\rm diff}, i/j} = \exp \biggl( -\frac{\Delta m_{i,j} g \Delta z}{k_{\rm B} T} \biggr) \simeq \exp \biggl[ -0.446 \biggl( \frac{\Delta z}{\rm 1\ km} \biggr) \biggl( \frac{T}{\rm 1\ K} \biggr)^{-1} \biggl( \frac{\Delta m_{i,j}}{\rm 1\ amu} \biggr) \biggr], \label{eq:Rdiff}
\end{equation}
where $k_{\rm B}$ is the Boltzmann constant, $g$ is the gravitational acceleration, $\Delta z$ is the distance between the homopause and exobase, $T$ is the temperature, and $\Delta m_{i,j}$ is the mass difference between two species. From Equation \ref{eq:Fsp}, the lifetime of species $i$ against sputtering loss, $\tau_i$, is given by,
\begin{equation}
    \frac{\tau_{i, {\rm sp}}}{\tau_{j, {\rm sp}}} = \frac{U_{{\rm es},i}}{U_{{\rm es},j}} R_{{\rm diff}, i/j}^{-1}. \label{eq:lifetime}
\end{equation}
We use Equation \ref{eq:lifetime} to relate the lifetime of Ne and other noble gases to that of Ar, which is estimated to be 4 $\times$ 10$^9$ years using the sputtering rate mentioned above. Importantly, the lifetime of minor species other than the dominant molecule CO$_2$ is independent from the abundance in atmosphere. This is because the loss rates of minor species are proportional to their abundances, and thus the dependence of lifetime on the abundance cancels out. 

The isotopic fractionation factor of sputtering is also given by $R_{{\rm diff}, i/j}$, as the difference in yields is expected to be small. In a steady state where the sputtering loss and supply is balanced, The isotope ratios of the atmosphere and the source are related by \citep{Jakosky+1994,Kurokawa+2018},
\begin{equation}
    \biggl( \frac{^{20}{\rm Ne}}{^{22}{\rm Ne}} \biggr)_{\rm source} = R_{{\rm diff}, ^{20}{\rm Ne}/^{22}{\rm Ne}} \cdot \biggl( \frac{^{20}{\rm Ne}}{^{22}{\rm Ne}} \biggr)_{\rm atm}. \label{eq:20Ne22Ne}
\end{equation}
Equation \ref{eq:20Ne22Ne} is used later in Section \ref{sec:origin} to discuss the origins of Martian Ne.

A comparison of Ne lifetime to that of atmospheric carbon (C) provides some insights into its behavior. A model combined to MAVEN's observations of precipitating ions estimated present-day the total C (CO$_2$ + CO + C) sputtering rate to be 3.4 $\times$ 10$^{23}$ s$^{-1}$ \citep{Leblanc+2019}. Dividing the total atmospheric CO$_2$ 3.3 $\times$ 10$^{41}$ molecules by the sputtering rate leads to $\tau_{\rm CO_2,sp}$ = 3 $\times$ 10$^{10}$ years. Because the dominant escape process of CO$_2$ (C) on present-day Mars is not sputtering but photodissociation and dissociative recombination \citep{Hu+2015,Lillis+2015}, the actual CO$_2$ lifetime is shorter and estimated to be 1 $\times$ 10$^{10}$ years (Figure \ref{fig:lifetime}). 

The lifetime of Ne in Mars' atmosphere is much shorter than C \citep{Jakosky+1994,Hutchins+1997}. This is caused by a lower mass of Ne, which leads to a higher scale height above the homopause and a higher yield (Equation \ref{eq:Fsp}). Recent measurements by MAVEN found that Mars' upper atmosphere is highly variable \citep[e.g.,][]{Jakosky+2017,Yoshida+2020}. Following \citet{Lammer+2020}, we assume $T$ = 200 K and $\Delta z$ = 60--80 km. The temperature was given by averaging the diurnal variations of the exospheric temperature \citep[127$\pm$8 to 260$\pm$7 K,][]{Stone+2018}. The homopause-exobase separation was given from the measurements of the homopause and exobase altitudes \citep{Jakosky+2017}. The adopted values result in $R_{{\rm diff}, {\rm Ne}/{\rm Ar}}$ = 15--35 (Equation \ref{eq:Rdiff}). Given $(x_{\rm Ne})_{\rm homo}$ = 1--6 ppm \citep{Owen+1977}, $(x_{\rm Ar})_{\rm homo}$ = 1.9\% \citep{Mahaffy+2013}, $U_{{\rm es,Ar}}/U_{{\rm es,Ne}}$ = 2, and $F_{\rm Ar,sp}$ = 4.9 $\times$ 10$^{22}$ s$^{-1}$ \citep{Leblanc+2019}, the sputtering loss rate of Ne is estimated to be $F_{\rm Ne,sp}$ = 0.7--10 $\times$ 10$^{20}$ s$^{-1}$ (Equation \ref{eq:Fsp}). The estimated lifetime of Ne $\tau_{\rm Ne,sp}$ is 0.6--1 $\times$ 10$^8$ years, while heavier noble gases have their lifetime comparable or longer than the age of Mars (Equation \ref{eq:lifetime} and Figure \ref{fig:lifetime}). 

We use the above estimated loss rate for atmospheric Ne to consider the balance between the loss and supply in the following sections. A background assumption here is that the sputtering rate estimated for current Mars represents the value averaged over a typical Ne lifetime ($\sim$ 10$^8$ years). A process which may influence the validity of this assumption is the variation of CO$_2$ partial pressure ($p_{\rm CO_2}$) due to the time-dependent obliquity \citep{Jakosky+1995,Nakamura+Tajika2003,Manning+2006}. A higher $p_{\rm CO_2}$ reduces Ne mixing ratio in the atmosphere and, consequently, its sputtering rate \citep{Jakosky+1994,Kurokawa+2018}. The chaotic nature of obliquity change inhibits its precise determination over more than 10 to 20 Myr \citep{Laskar+2004}. Considering that deposits of CO$_2$ ice that are buried in the south-polar region today \citep[equivalent to $p_{\rm CO_2}$ = 600 Pa,][]{Phillips+2011,Bierson+2016,Jakosky2019} could have been in the atmosphere during periods of higher obliquity, the upper limit of $p_{\rm CO_2}$ is twice of the current value. Thus, our estimate for Ne loss rate from Mars in the recent $\sim$ 10$^8$ years may be overestimated by a factor of two in maximum, which is smaller than the uncertainty in Ne loss rate from current Mars and thus acceptable for the accuracy of our estimates given below.

We note that, while we are interested in the loss rates of atmospheric species for recent Mars, these were likely higher for early Mars due to the more active Sun in billions of years ago \citep[e.g.,][]{Wood+2002,Ribas+2005,Tu+2015}, as recorded in the enrichment of heavy isotopes \citep[e.g.,][]{Jakosky+1994,Jakosky+2017,Jakosky+2018,Hu+2015,Slipski+Jakosky2016,Kurokawa+2018,Lammer+2020}. The short lifetime of Ne (0.6--1 $\times$ 10$^8$ years) compared to the timescale of solar evolution ($\sim$10$^9$ years) justifies our treatment.

\subsection{Supply by asteroids, comets, and interplanetary dust particles}
\label{subsec:asteroids}

Without continuous supply, atmospheric Ne abundance should decrease in time with the $e$-folding time identical to the lifetime estimated in Section \ref{subsec:escape}. This suggests that Ne had been supplied to the atmosphere until recent (<$\tau_{\rm Ne,sp}$), or it is being supplied even on present-day Mars. As Ne sputtering loss rate is proportional to atmospheric Ne abundance (Equation \ref{eq:Fsp}), continuous supply of Ne leads to its balance with sputtering loss \citep{Kurokawa+2018}. Assuming the balance between the estimated sputtering rate (Section \ref{subsec:escape}) and continuous supply from any source simply lead to the supply rate $F_{\rm Ne, sup}$ = $F_{\rm Ne, sp}$ = 0.7--10 $\times$ 10$^{20}$ s$^{-1}$. Assuming the supply cessation $\tau_{\rm cess}$ ago result in the supply rate $e^{\tau_{\rm cess}/\tau_{\rm Ne,sp}}$ times higher in the past, as the current Ne abundance should have evolved from a higher level.

The estimated lifetime $\tau_{\rm Ne,sp}$ and abundance of atmospheric Ne rule out recent asteroid and comet impacts and interplanetary dust particles (IDPs) accretion as its origins (Figure \ref{fig:overview}). \citet{Frantseva+2018} estimated mass flux of C-type asteroids and comets onto Mars as 2.5 $\times$ 10$^6$ kg/yr and 0.13 $\times$ 10$^6$ kg/yr, respectively. Ne abundance of carbonaceous chondrites is 4 $\times$ 10$^{-10}$ g/g \citep{Marty2012}. Ne was not detected by Rosetta's in-situ observations of comet 67P/Churyumov-Gerasimenko, which gives an upper limit on Ne abundance as <3 $\times$ 10$^{-8}$ g/g \citep{Rubin+2018}. Using these estimates yield Ne supply flux by C-type asteroids and comets as 1 $\times$ 10$^{15}$ s$^{-1}$ and <5 $\times$ 10$^{15}$ s$^{-1}$, respectively. We note that the contribution of S-type asteroids is negligible, given the lower Ne abundance of ordinary chondrites \citep{Busemann+2000}. IDPs' high Ne abundance due to the solar-wind implantation provides higher Ne supply flux, but it is still limited to 5 $\times$ 10$^{18}$ s$^{-1}$ \citep{Flynn+1997}, which is lower than Ne sputtering flux by more than an order of magnitude. We note that the estimate of dust flux onto Mars varies among studies \citep{Flynn+1997,Borin+2017,Crismani+2017,Frantseva+2018}, and here we adopted the highest value. The release of trapped Ne from IDPs requires a heating process after accretion. Thus, the value adopted above, which assumed instantaneous release, is an upper limit of IDPs' contribution.

\subsection{Supply of cosmogenic Ne}
\label{subsec:cosmogenic}

Degassing of cosmogenic Ne from surface rocks cannot be a major source of atmospheric Ne as well (Figure \ref{fig:overview}). Due to the thin atmosphere of Mars, the surface rocks are exposed to galactic cosmic rays, which causes Ne production from spallation of Mg, Si, and Al \citep{Farley+2014}. Such cosmogenic contribution has been proposed for Kr and Xe in the Martian atmosphere \citep{Conrad+2016}. In the case of Ne, however, the production rate is found to be only $\sim$1 $\times$ 10$^{18}$ s$^{-1}$. Here we adopted a uniform production rate $^{21}$Ne 0.02 pmol/g/Ma for the topmost 2 m of the crust \citep{Farley+2014} and $^{20}$Ne:$^{21}$Ne:$^{22}$Ne to be $\sim$1:1:1 \citep{Lal1993}. Moreover, the measured $^{36}$Ar/$^3$He and $^{36}$Ar/$^{21}$Ne at Curiosity's drilling sites are consistent with negligible release of the gases produced in the rock \citep{Farley+2014,Vasconcelos+2016}. 

\subsection{Volcanic degassing and mantle Ne content}
\label{subsec:mantleNe}

Given the limited contribution of the other sources (Sections \ref{subsec:asteroids} and \ref{subsec:cosmogenic}), volcanic degassing is the most promising origin of atmospheric Ne (Figure \ref{fig:overview}). Here we estimate the lower limit of Ne abundance in the current Martian mantle which can sustain Ne volcanic degassing rate in balance with the sputtering loss to space.

Though the major volcanic activity is limited to the late Noachian and Hesperian, high-resolution crater-based studies have identified recent ($\sim10^{7-8}$ years) caldera floors and lava flows in the Tharsis region \citep{Hauber+2011,Robbins+2011,Grott+2013}. After the submission of this study, evidence of very recent volcanic activity as young as 5 $\times$ 10$^4$ years in Elysium Planitia was reported \citep{Horvath+2021}. These studies imply that a certain level of volcanic activity persisted up to present day. The magma eruption rate of those recent activity has not been estimated, and here we adopt that for the late Amazonian as an upper limit. Dividing the maximum estimate for total (extrusive and intrusive) magma volume 2.7 $\times$ 10$^7$ km$^3$ \citep{Greeley+Schneid1991} by the duration of the late Amazonian 0.3--0.6 Gyrs \citep{Hartmann+Neukum2001} leads to 0.045--0.09 km$^3$/yr, and thus the upper limit is 0.09 km$^3$/yr. Here we assumed the highest estimate of intrusive-to-extrusive eruption ratio, 12:1 \citep{Greeley+Schneid1991}. 
Adopting a typical magma density 3.3 g/cm$^3$ and complete degassing even from intrusive magma, Ne abundance in the source magma (the magma chamber in Figure \ref{fig:overview}) to balance volcanic supply of Ne with the sputtering loss is estimated to be >0.3--4 $\times$ 10$^{-9}$ g/g, where the range comes from the model uncertainty in Ne loss rate (Section \ref{subsec:escape}). 

The source magma is produced by partial melting of mantle upwelling. An upwelling mantle plume experiences partial melting \citep{Li+Kiefer2007}, forming many small melt particles within it. Each small melt particle is equilibrated with the surrounding residual rocks. Owing to its high incompatibility, Ne is almost fully partitioned into the melt \citep{Heber+2007,Moreira+Kurz2013}. Melt is channeled and coalesces to form a dike, which sources magma to a shallow-level magma chamber or directly erupts into the surface \citep{ONeil2007}. The melt fraction of the source mantle is estimated to be 0.02--0.1 from the analysis of rare earth elements in Martian basaltic meteorites \citep{Norman1999,Borg+Draper2003}. Thus, multiplying the above estimated Ne abundance in the magma by the lower limit of the melt fraction 0.02 ends up with Ne abundance of the current mantle $x_{\rm Ne, mantle}^{\rm present}$ >0.5--8 $\times$ 10$^{-11}$ g/g (Figures \ref{fig:overview} and \ref{fig:abundances}). 

\section{Ne content in the magma ocean stage}
\label{sec:mantle}

\begin{figure}
    \centering
    \includegraphics[width=\linewidth]{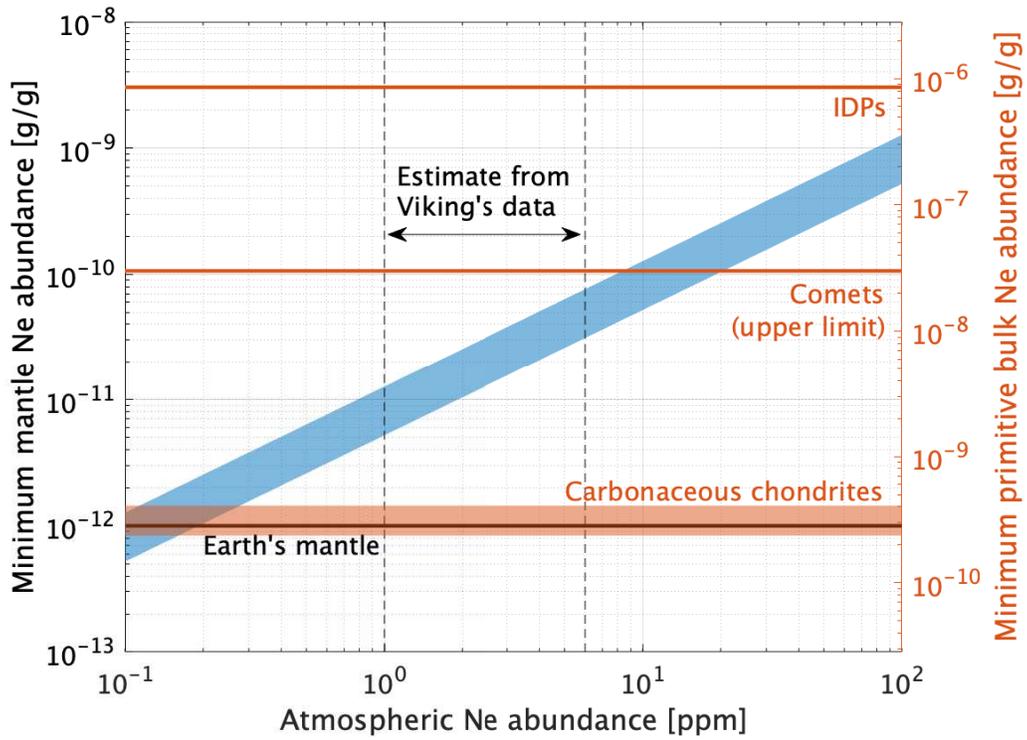}
    \caption{Minimum estimates for mantle (left axis, Section \ref{sec:atmosphere}) and primitive bulk (right axis, Section \ref{sec:mantle}) Ne abundances as a function of that of atmospheric Ne (blue area). The balance between the atmospheric loss and volcanic supply is assumed. The range of blue area comes from the uncertainty in the elemental fractionation factor of sputtering. Dashed lines denote the range of atmospheric Ne abundance estimated from Viking data \citep{Owen+1977}. Earth's bulk mantle Ne abundance \citep{Marty2012} is shown for comparison to the Martian mantle value (left axis). Orange lines are Ne abundances in possible sources to compare with the Martian primitive bulk value (right axis). Carbonaceous chondrites: \citet{Marty2012}. Comets: \citet{Rubin+2018}. IDPs: \citet{Flynn+1997}.}
    \label{fig:abundances}
\end{figure}

The Martian mantle is rich in Ne. In Section \ref{subsec:mantleNe}, we estimated the abundance $x_{\rm Ne, mantle}^{\rm present}$ to be >0.5--8 $\times$ 10$^{-11}$ g/g. The total Ne mass in the mantle $M_{\rm Ne, mantle}^{\rm present}$ is >3--40 $\times$ 10$^{12}$ kg \citep[assuming the mantle mass fraction 0.76,][]{Rivoldini+2011}. As the total Ne mass in the atmosphere is 1--6 $\times$ 10$^{10}$ kg, this estimate means that the mantle is possibly the largest reservoir of Ne on Mars. For comparison, Earth's largest reservoir of Ne is its atmosphere, and Ne abundance in Earth's present-day bulk mantle has been estimated to be 1 $\times$ 10$^{-12}$ g/g from calibration to $^{40}$Ar \citep{Marty2012}. Thus, Ne abundance in the Martian mantle is higher than in Earth's by more than a factor of 5--80. We note that here we compared their present-day abundances, and the relation may not be applicable to their original values (at the time when their magma oceans solidified) because the mantle Ne abundances would have declined with time due to secular degassing (see also Section \ref{subsec:EarthMars}). 

The high Ne abundance suggests that the Martian mantle is rich in other volatile elements as well (Section \ref{sec:origin}) and that secular degassing from such a volatile-rich mantle could have been an important driver of Mars' climate. Moreover, as we explain below, the smaller size of Mars relative to Earth requires an even larger amount of total Ne in the magma ocean stage to put Ne into the mantle.

The abundant Ne in the Martian mantle likely have its origin in the planet formation stage. Plate tectonics and crustal recycling have not been operating on Mars at least for the past 4 billion years \citep[e.g.,][]{Grott+2013}. Subduction-driven Ne recycling is inefficient even on tectonically-active Earth \citep{Bekaert+2020}. The solar-like $^{20}$Ne/$^{22}$Ne ratio of the deep mantle endmember \citep[13.23$\pm$0.22,][]{Williams+Mukhopadhyay2019} is distinct from that in Earth's atmosphere (9.8) and suggests that Earth's mantle Ne originated at least partially from captured proto-solar nebula gas \citep{Yokochi+Marty2004}.

Incorporating the large amount of Ne into the mantle requires an even larger amount of total Ne during the formation stage. This is because of Ne's low solubility in silicate melt and incompatibility. Here we consider Ne partitioning between the atmosphere, magma ocean, and solidified mantle in the magma ocean stage. The partial pressure of Ne in the atmosphere in equilibrium with the magma ocean $P_{\rm Ne}$ is given by,
\begin{equation}
    P_{\rm Ne} = \frac{x_{\rm Ne, mo}}{S_{\rm Ne}}, \label{eq:PNe}
\end{equation}
where $x_{\rm Ne, mo}$ is Ne mass fraction in the magma ocean and $S_{\rm Ne}$ is the solubility. The total Ne mass in the atmosphere $M_{\rm Ne, atm}$ is given by,
\begin{equation}
    M_{\rm Ne, atm} = \frac{m_{\rm Ne}}{\overline{m}} \frac{A p_{\rm Ne}}{g}, \label{eq:MNeatm}
\end{equation}
where $\overline{m}$ is the mean molecular mass of atmospheric gases and $A$ is the surface area of the planet. Given Equations \ref{eq:PNe}, \ref{eq:MNeatm}, and $M_{\rm Ne, mo} = x_{\rm Ne, mo} \cdot M_{\rm mo}$ (where $M_{\rm mo}$ is the mass of magma ocean), the atmospheric and magma ocean Ne mass ratio is given by,
\begin{equation}
    \frac{M_{\rm Ne, atm}}{M_{\rm Ne, mo}} = \frac{m_{\rm Ne} A}{\overline{m} g S_{\rm Ne} M_{\rm mo}}.
\end{equation}
Assuming a pyrolite magma composition \citep{Mcdonough+Sun1995} and the temperature of 1,700 K in a semi-empirical solubility model of \citet{Iacono-Marziano+2010}, the solubility $S_{\rm Ne}$ is estimated to be 1.6 ppm/MPa, which leads to $M_{\rm Ne, atm}/M_{\rm Ne, mo}$ > 20, where the minimum value is given by assuming a fully molten magma ocean ($M_{\rm mo}$ = $M_{\rm mantle}$, where $M_{\rm mantle}$ is the mass of the mantle). 

Moreover, the mineral-melt partitioning coefficient of Ne is extremely low \citep[$\sim$10$^{-4}$,][]{Heber+2007}, which means that magma ocean solidification induces significant degassing. Thus, the mass of Ne trapped in the solidified mantle would be dominated by Ne in interstitial melt rather than in crystallized minerals. Given the trapped melt fraction $F_{\rm tl}$, here defined as the mass fraction of trapped interstitial melt relative to the total solidified mass (trapped and then solidified melt plus crystallized minerals) at the solidification front, Ne mass in the solidified mantle $M_{\rm Ne, sm}$ upon magma ocean solidification is given by,
\begin{equation}
    dM_{\rm Ne, sm} = F_{\rm tl} \cdot x_{\rm Ne, mo} \cdot dM_{\rm sm}. \label{eq:dMNe}
\end{equation}
We note that the subscript sm is used for the solidified mantle under the magma ocean, whereas the subscript mantle is for the mantle after complete solidification.
Assuming $F_{\rm tl}$ to be constant through the bottom-up solidification, Equation \ref{eq:dMNe} can be integrated analytically. The mass of Ne in the mantle after complete mantle solidification is given by (see \ref{ap:MNe} for derivation),
\begin{equation}
    M_{\rm Ne, mantle} = \biggl[ 1- \biggl( \frac{1}{1 + \overline{m} g S_{\rm Ne} M_{\rm mantle}/m_{\rm Ne} A } \biggr)^{F_{\rm tl}} \biggr] M_{\rm Ne, tot}, \label{eq:MNemantle}
\end{equation}
where $M_{\rm Ne, tot}$ is the total mass of Ne in the system. In the low solubility limit ($\overline{m} g S_{\rm Ne} M_{\rm mantle}/m_{\rm Ne} A \ll 1$, which is satisfied for Ne), Equation \ref{eq:MNemantle} is approximated by,
\begin{equation}
    M_{\rm Ne, mantle} \simeq F_{\rm tl} \frac{\overline{m} g S_{\rm Ne} M_{\rm mantle}}{m_{\rm Ne} A} M_{\rm Ne,tot}. \label{eq:Mnemantleapp}
\end{equation}
The trapped melt fraction is typically assumed to be $\sim$1\%, but it would be dependent on the evolutionary properties of magma ocean \citep[e.g., the cooling timescale,][]{Hier-Majumder+Hirschmann2017}. Assuming the maximum estimate $F_{\rm tl} = 0.3$, which corresponds to the value for the rheological transition, we obtain $M_{\rm Ne, tot} > 3\times 10^{2} \ M_{\rm Ne, mantle}$ for Mars. Thus, assuming that $M_{\rm Ne, mantle} = M_{\rm Ne, mantle}^{\rm present}$, >300 times more Ne is required to put the estimated Ne abundance into the Martian mantle (Figure \ref{fig:abundances}). We note that such preferential partitioning into the atmosphere in the magma ocean stage has been modeled also for C \citep{Elkins2008,Hier-Majumder+Hirschmann2017}. The excess amount of Ne in the atmosphere would have been lost after the magma ocean solidification by atmospheric escape processes such as impact erosion, hydrodynamic escape, and sputtering.

Secular degassing after the magma ocean solidification requires even higher primitive Ne content ($M_{\rm Ne, mantle} > M_{\rm Ne, mantle}^{\rm present}$). Integrating the estimated volcanic degassing rate (Section \ref{subsec:mantleNe}) of Ne on current Mars for 4.5 Gyrs leads to $\simeq$10\% depletion of the estimated mantle Ne content. While the degree of cumulative volcanic degassing cannot be directly constrained and is dependent on models \citep{Grott+2013}, higher volcanic activity of early Mars \citep{Greeley+Schneid1991} would cause more depletion. Thus, the estimated primitive Ne content ($M_{\rm Ne, tot} > 3\times 10^{2} \ M_{\rm Ne, mantle}$) is the lower limit.

Another useful measure of the required Ne mass is the partial pressure. Combining Equation \ref{eq:PNe} and \ref{eq:Mnemantleapp}, we obtain,
\begin{equation}
    p_{\rm Ne} \simeq \frac{x_{\rm Ne,mantle}}{F_{\rm tl} S_{\rm Ne}}. \label{eq:PNerequired}
\end{equation}
Substituting $F_{\rm tl}<0.3$ and $x_{\rm Ne,mantle}>x_{\rm Ne,mantle}^{\rm present}$ for Equation \ref{eq:PNerequired} gives the required Ne partial pressure of the Martian primitive atmosphere overlying the magma ocean to be > 10 Pa. This required Ne partial pressure corresponds to 300 times the current mantle Ne content derived above. The pressure is compared with that of the captured solar nebula gas in Section \ref{sec:origin}.

Equation \ref{eq:Mnemantleapp} exhibits the dependence on the planet size $M_{\rm Ne, tot}/M_{\rm Ne, mantle} \propto M_{\rm p}^{-\frac{2}{3}} \overline{\rho}^{-\frac{4}{3}}$, where $M_{\rm p}$ and $\overline{\rho}$ are the mass and the bulk density of the planet, respectively. The right-hand side factor is $\sim$7 times larger for Mars than for Earth, due to the smaller gravity and the higher surface area to mass ratio. This means that putting the same amount of Ne into the Martian mantle requires a total amount of Ne 7 times larger than that for Earth.

\section{Origin of abundant Martian Ne}
\label{sec:origin}

\begin{figure}
    \centering
    \includegraphics[width=\linewidth]{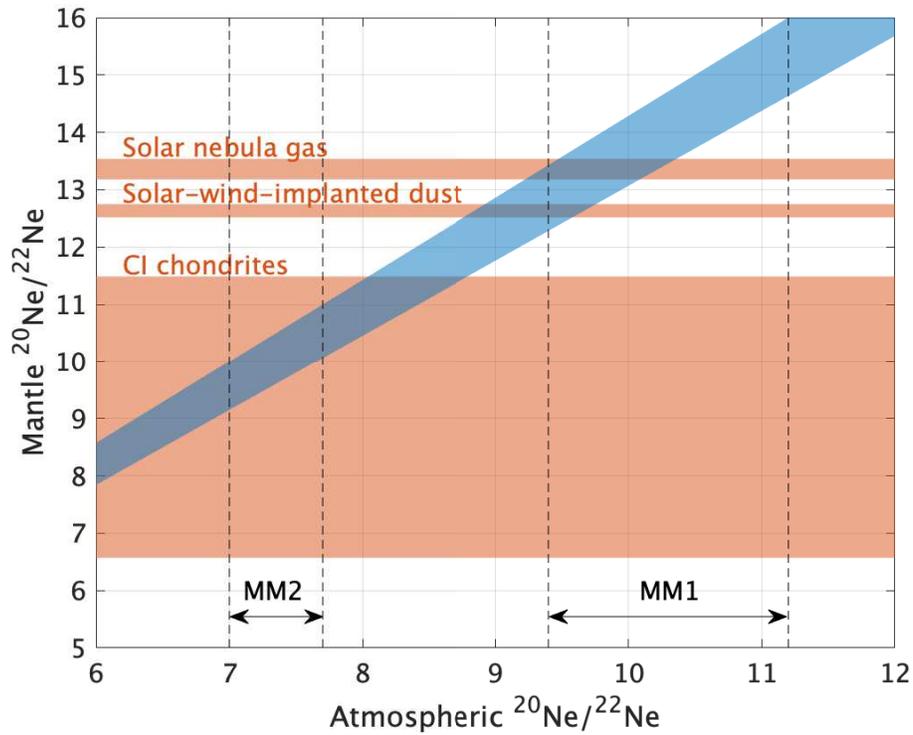}
    \caption{$^{20}$Ne/$^{22}$Ne ratios in the Martian atmosphere and mantle. The blue area denotes the steady-state relation where the atmospheric loss and supply are balanced (Equation \ref{eq:20Ne22Ne}). Its range comes from the uncertainty in the isotopic fractionation factor of sputtering. The ranges of atmospheric values suggested from trapped gas in Martian meteorites are indicated by vertical lines. MM1: \citet{Wiens+1986,Swindle+1986}. MM2: \citet{Park+Nagao2006,Park+2017}. The isotope ratios of possible sources are shown to compare with the mantle values (orange areas). Solar nebula gas: \citet{Heber+2012}. Solar-wind-implanted dust: \citet{Moreira+2016,Peron+2017}. CI chondrites: \citet{Mazor+1970}.}
    \label{fig:isotopes}
\end{figure}

Here we combine the isotopic constraint on Ne source (Section \ref{subsec:knowledge}) and the estimates of the required Ne content and atmospheric partial pressure in the magma ocean stage (Section \ref{sec:mantle}) to discuss the possible origins of abundant Martian Ne.

Atmospheric $^{20}$Ne/$^{22}$Ne ratios informed from Martian meteorites (Section \ref{subsec:knowledge}) are consistent with either the solar nebula gas, the solar-wind implanted dust, or the chondritic origin, yet the uncertainty in the true atmospheric value prevents the conclusive determination of the origin(s) (Figure \ref{fig:isotopes}). If the atmospheric $^{20}$Ne/$^{22}$Ne ratio is $\sim$10 (MM1 in Figure \ref{fig:isotopes}), the estimated mantle $^{20}$Ne/$^{22}$Ne ratio suggests the captured solar nebula or the solar-wind-implanted dust origin. If the atmospheric $^{20}$Ne/$^{22}$Ne ratio is $\sim$7 (MM2 in Figure \ref{fig:isotopes}), The estimated mantle $^{20}$Ne/$^{22}$Ne ratio suggests the chondritic origin.

Prior to future precise determination, below we discuss the implications for the abundance of volatile elements in Mars and for accretion history of Mars from the three possible Ne origins.

\subsection{Capture of solar nebula gas}
\label{subsec:capture}

A planet formed in a protoplanetary disk captures disk gas as the atmosphere (hereafter called the primordial atmosphere) by its gravity \citep[e.g.,][]{Hayashi+Nakazawa1979,Sasaki1999}. Its $^{20}$Ne/$^{22}$Ne ratio should be identical to the solar value (13.36$\pm$0.16, Figure \ref{fig:isotopes}). The primordial atmospheres are not present on the terrestrial planets in our Solar system. However, the $^{20}$Ne/$^{22}$Ne ratio of Earth's deep mantle Ne is consistent with the solar nebula origin \citep{Yokochi+Marty2004,Williams+Mukhopadhyay2019}. Earth's water as well as hydrogen (H) in the core has been proposed to originate at least partially from H in the primordial atmosphere \citep{Genda+Ikoma2008,Hallis+2015,Sharp2017,Wu+2018,Olson+Sharp2018,Olson+Sharp2019,Saito+Kuramoto2020}. A low deuterium to hydrogen (D/H) ratio signature of Martian meteorite has been also attributed to the nebula-gas origin \citep{Hallis+2012,Hallis+2015}, though a higher, chondrite-like D/H ratio is considered as a typical mantle value \citep{Usui+2012}. A solar-like Xe isotopic composition found in Chassigny, which is interpreted to originate from the Martian interior \citep{Ott1988}, may also support capture of the solar nebula gas. The elemental abundance pattern of Venus' atmospheric noble gases suggests the remnant of the primordial component \citep{Pepin1991,Genda+Abe2005}. The primordial atmospheres are ubiquitous in extrasolar planetary systems, known as super Earths and mini-Neptunes \citep[e.g.,][]{Rogers2015}.

While forming a dense primordial atmosphere is relatively easy for an Earth-sized planet, it is possible for a Mars-sized planet only under some specific conditions as we show below. This is because the amount of the captured primordial atmosphere highly depends on the planetary mass \citep{Sasaki+Nakazawa1990,Ikoma+Genda2006}. Therefore, the solar-nebula-gas originated Ne in the Martian mantle can, if proved, serve important constraints on its formation history. 

Assuming a spherically-symmetric, radiative-equilibrium structure \citep{Sasaki+Nakazawa1990}, the surface pressure of the primordial atmosphere surrounding a planet embedded in a protoplanetary disk is approximated by,
\begin{equation}
    p_{\rm s} \simeq \frac{\pi^2\sigma}{36} \biggl( \frac{G \mu m_{\rm H}}{k_{\rm B}} \biggr)^4 \overline{\rho} M_{\rm p}^2 \kappa^{-1} \tau_{\rm acc}, \label{eq:ps}
\end{equation}
where $k_{\rm B}$ is the Boltzmann constant, $\sigma$ is the Stefan–Boltzmann constant, $G$ is the gravitational constant, $m_{\rm H}$ is the mass of a H atom, $\mu$ is the mean molecular weight, $\kappa$ is the opacity, and $\tau_{\rm acc}$ is the accretion time scale which gives the luminosity $L = GM_{\rm p}^2/R_{\rm p}\tau_{\rm acc}$. We note that Equation \ref{eq:ps} has planetary-mass dependence different from that in Equation 15 of \citet{Sasaki+Nakazawa1990} (the power-law index 2 vs. 3), the latter of which seems to be a typo. Substituting Mars mass and density into Equation \ref{eq:ps} leads to,
\begin{equation}
    p_{\rm s} \simeq 1\times 10^{3}\ \biggl( \frac{\mu}{2.35} \biggr)^4 \biggl( \frac{\kappa}{0.1\ {\rm m^2/kg}} \biggr)^{-1} \biggl( \frac{\tau_{\rm acc}}{1\ {\rm Myr}}  \biggr) \ {\rm Pa}, \label{eq:ps_scaled}
\end{equation}
where we adopted the mean molecular weight of solar nebula gas, an interstellar-medium-like opacity \citep[e.g.,][]{Stevenson1982}, and an estimated growth timescale for Mars \citep[1--3 Myrs,][]{Dauphas+Pourmand2011} as the scaling factors. 
%We note that the derivation of Equation \ref{eq:ps_scaled} have already assumed a dense atmosphere (the surface pressure and temperature are much higher than  those of the surrounding solar nebula), and thus it is not good approximation under the conditions of those scaling factors. The assumption is justified as we consider conditions to form a dense atmosphere below. 
Substituting the molar fraction of $^{20}$Ne in the solar nebula $f_{\rm Ne,SN}$ = 1.8$\times$10$^{-4}$ \citep{Wieler+2002,Yokochi+Marty2004} into Equation \ref{eq:ps_scaled}, we obtain the Ne partial pressure as,
\begin{equation}
    p_{\rm Ne} \simeq 2\times 10^{-1} \biggl( \frac{f_{\rm Ne}}{f_{\rm Ne,SN}} \biggr) \biggl( \frac{\mu}{2.35} \biggr)^4 \biggl( \frac{\kappa}{0.1\ {\rm m^2/kg}} \biggr)^{-1} \biggl( \frac{\tau_{\rm acc}}{1\ {\rm Myr}}  \biggr) \ {\rm Pa}. \label{pNe_scaled}
\end{equation}
Equation \ref{pNe_scaled} shows that Ne partial pressure of the Martian primordial atmosphere under a typical condition is two orders of magnitude lower than the value required to put the estimated Ne into the mantle ($\sim$ 10 Pa, Section \ref{sec:mantle}).

What are the formation scenarios of Mars which enable the accretion of a dense primordial atmosphere? A possibility is the completion of accretion prior to the dissipation of the solar nebula, which lowers the accretion luminosity (long $\tau_{\rm acc}$ in Equation \ref{pNe_scaled} at the late stage). This scenario may be consistent with Mars accretion time estimated from hafnium-tungsten-thorium (Hf–W–Th) chronology \citep{Dauphas+Pourmand2011} being earlier than the dissipation time of the solar nebula suggested from meteorite paleomagnetism \citep{Weiss+2021}. The opacity $\kappa$, which is mainly determined by the dust abundance, can be also lowered, if the surrounding solar nebula at the time of gas dissipation was depleted in small dust. Both of the two possibilities point to a low surface density of solid materials in Mars forming region, which provides constraints on dust coagulation and transport processes in the solar nebula.

Another possibility to increase Ne partial pressure is mixing of a high-molecular-weight component degassed from Mars building blocks. For instance, mixing degassed CO$_2$ (other components are ignored for simplification) and the captured solar nebula gas with 1:1 ratio leads to $f_{\rm Ne}/f_{\rm Ne,SN}$ = 0.5 and $\mu/2.35 \sim$ 20, and thus $p_{\rm Ne} \sim$ 1$\times$10$^4$ Pa (Equation \ref{eq:ps_scaled}). Though \citet{Saito+Kuramoto2018,Saito+Kuramoto2020} concluded that the two components are likely stratified rathen than mixed, eddy diffusion or mechanical mixing either due to impacts or nebula gas flow \citep{Ormel+2015,Kurokawa+Tanigawa2018,Kuwahara+2019,Mai+2020} may induce the mixing. 

If the mantle Ne resulted from the nebula gas capture, the mantle would acquire other volatile elements from the nebula gas depending on their solubility into the magma ocean, partitioning coefficients between minerals and silicate melt, and those between core-forming metals and silicate melt. The solubilities of major volatile elements H (which forms water), N, and S \citep[e.g.,][]{Hirschmann2016} are higher than that of Ne \citep{Iacono-Marziano+2010}. This means that these elements can be trapped from solar nebula gas more efficiently than Ne. 

\subsection{Solar-wind-implanted dust}

Exposure of small dust grains to the solar wind implants its ions to their surfaces. A combination of the solar-wind implantation and erosion of the outer layers of the grains, where $^{20}$Ne enriches, results in the $^{20}$Ne/$^{22}$Ne ratio slightly lower than the solar value (12.52--12.75, Figure \ref{fig:isotopes}). The solar-wind-implanted dust has been proposed to the origin of Earth's mantle Ne alternative to the nebula gas \citep{Trieloff+2000,Moreira+2016,Jaupart+2017,Peron+2017,Vogt+2019}. We note that both solar-wind-implanted dust and IDPs are the same materials, but the former and latter denote those in the planet formation stage and in the current Solar System, respectively, in our definition following the terminology in the literature.

The solar-wind implanted dust grains could be rich in Ne, as known for IDPs \citep{Flynn+1997}. Assuming Ne abundance of IDPs ($\sim$9$\times$10$^{-7}$ g/g), the estimated primordial bulk abundance of Ne can be supplied if Mars building blocks are consist of $\sim$0.1--1 wt.\% irradiated materials (Figure \ref{fig:abundances}). The actual Ne abundances of these grains depend on their grain sizes, distances from the Sun, and exposure time \citep{Moreira+2016}.

If Martian Ne originated from the solar-wind-implanted dust, it provides constraints on dust grain sizes and turbulent strengths in the solar nebula when Mars formed, both of which are important parameters for the planet formation \citep[e.g.,][]{Ormel+Klahr2010,Lambrechts+Johansen2014,Guillot+2014,Kuwahara+Kurokawa2020a,Kuwahara+Kurokawa2020b}. Mars likely formed before the dissipation of the solar nebula \citep{Dauphas+Pourmand2011,Kobayashi+Dauphas2013}. Thus, Mars forming region was protected from the solar wind irradiation by the solar nebula. The exposure of Martian building materials to the solar wind requires dust grains being small enough and turbulence in the solar nebula to be strong enough in order to dredge up them from the midplane to higher altitudes of the solar nebula disk \citep{Moreira+2016}. Grain sizes and turbulent strength in protoplanetary disks have been studied actively with telescopic observations \citep{Kataoka+2016,Dullemond+2018,Okuzumi+Tazaki2019,Ohashi+Kataoka2019,Rosotti+2020,Ueda+2020,Doi+Kataoka2021}, but these studies focus on outer regions of protoplanetary disks as the spatial resolution of observations is typically limited to several tens to hundreds au. Thus, the presence of the solar-wind Ne component in the Martian mantle is, if proved, an important clue to understand the property of inner, terrestrial planet forming regions of protoplanetary disks ($\sim$ 1 au). 

Alternatively, formation of Mars after the nebula dissipation may be responsible for the accretion of the solar-wind-implanted dust. \citet{Marchi+2020} proposed that impact generation of mantle domains with variably fractionated Hf/W ratios and diverse $^{182}$W can relax the constraint on the formation timescale \citep{Dauphas+Pourmand2011}. 
%\citet{Moreira+2016} showed that efficient irradiation is possible by assuming strong turbulence (the turbulence parameter 0.01) 

Solar wind implantation is in principle applicable to any types of materials. Therefore, the amounts of other volatile elements supplied to accreting Mars are not directly constrained in this scenario, though the solar-wind-implanted bodies are in general rich in volatile elements (see Section \ref{subsec:chondtites_and_comets}). Additionally, volatile elements other than Ne (e.g., H) can be implanted to Mars building blocks and supplied to Mars as well.

% DoiKataoka, OhhashiKataoka, Dullmond, Rosotti, Ueda?, kataoka16, OkuzumiTazaki2019

\subsection{Chondrites and comets}
\label{subsec:chondtites_and_comets}

Because chondrites are thought to be fundamental building blocks of terrestrial planets, their contribution as Ne source is naturally expected. Carbonaceous chondrites show $^{20}$Ne/$^{22}$Ne = 9.03$\pm$2.46 (Figure \ref{fig:isotopes}) as a non-cosmogenic endmember. The origin of this endmember has been proposed to be a mixture of the presolar component and the other (so called Q-gas or the solar wind) component \citep[][and references therein]{Moreira+2016}. Ordinary and enstatite chondrites show broadly similar $^{20}$Ne/$^{22}$Ne ratios, but the contribution of the solar component seems to be higher \citep{Moreira+2016}.

However, even considering the most volatile-rich, carbonaceous chondrites does not provide enough Ne to explain the estimated Martian mantle Ne content (Figure \ref{fig:abundances}). Therefore, if the Martian mantle is proved to have a chondritic $^{20}$Ne/$^{22}$Ne ratio, it suggests unknown Martian building blocks that are rich in Ne with a chondritic isotope composition. This is also related to the elemental composition of bulk Mars -- for instance, Ne in carbonaceous chondrites is trapped in carbon phases \citep[e.g.,][]{Busemann+2000} and thus a hypothetical Ne-rich materials may be rich in C as well.

Comets are potential candidates of such Ne-rich bodies which might constitute some fraction of Martian building blocks. In situ measurements of gases in the coma of comet 67P/Churyumov-Gerasimenko did not detect Ne \citep{Rubin+2018}, and thus only the upper limit on Ne abundance was provided (Figure \ref{fig:abundances}). Ne trapped in ice is thought to have a solar-like isotopic composition \citep{Dauphas2003}. In contrast, refractory grains returned by the Stardust mission to comet Wild 2 show rather chondritic Ne isotopic compositions with their abundances comparable to IDPs \citep{Marty+2008,Palma+2019}. The determination of the bulk abundance and isotopic composition of Ne in comets would thus be important to understand the origin of Martian Ne.

In both cases, Ne is relatively depleted in chondrites and possibly in comets compared to the solar abundance \citep[e.g.,][]{Dauphas2003,Marty2012}. Thus, the Ne-rich mantle (Section \ref{subsec:mantleNe}) suggests enrichment of other volatile elements as well.

\section{Discussion}
\label{sec:discussion}

\subsection{Measurements of atmospheric Ne by future missions}

Given its importance for constraining the abundance and origin of Ne in the mantle (Section \ref{sec:origin}), more precise determination of the elemental and isotopic abundances of Ne in the Martian atmosphere is desired. Though indirect determination using trapped gas in Martian meteorites has provided useful estimates (Section \ref{subsec:knowledge}), in situ measurements would, if performed, provide undoubted constraints. Here we discuss the feasibility of in situ measurements by a future exploration mission. As mentioned in Section \ref{subsec:knowledge}, only the abundance of $^{22}$Ne has been determined for Ne with in situ analysis. The most abundant isotope $^{20}$Ne has not been measured so far due to the interference of $^{40}$Ar$^{++}$ to $^{20}$Ne$^{+}$ in mass spectrometry. The abundance of $^{40}$Ar is much higher than that of $^{20}$Ne in the atmosphere \citep[2.5 ppm for $^{20}$Ne and 2\% for $^{40}$Ar, respectively;][]{Owen+1977,Franz+2017}. Therefore, Ar must be removed for the Ne isotope measurements. Although Ar is separated cryogenically by using liquid nitrogen and/or a cryo-trap in laboratory experiments, it is difficult to apply this method to space missions. As a preferable way to separate Ne from Ar, \citet{Miura+2020} proposed a method using a permeable membrane. They experimentally investigated permeation of Ne and Ar through polyimide and Viton sheets under the terrestrial atmosphere, where polyimide sheets more efficiently separated Ne from Ar.

Determining whether the Martian mantle Ne is solar-like (solar nebula or implanted solar wind, $^{20}$Ne/$^{22}$Ne $\sim$ 13) or chondritic ($^{20}$Ne/$^{22}$Ne $\sim$ 7--11) requires the measurements of the atmospheric Ne isotopic ratio with $<$10\% uncertainty (Figure \ref{fig:isotopes}). On the basis of the experimental data, a polyimide sheet with a thickness of $\sim$75 $\mu$m and an area of $\gtrsim$50 cm$^{2}$ has been suggested as a possible membrane for the Ne-Ar separation under the Martian atmospheric condition \citep{Miura+2020}. Within a reasonable time (e.g., within an hour), a detectable amount of Ne does not permeate through a polyimide sheet much thicker than 75 $\mu$m, on the other hand, a significant amount Ar as well as Ne permeates through the sheet thinner than 75 $\mu$m. The necessary size of the membrane depends on the sensitivity of the mass spectrometer. The amount of permeating $^{20}$Ne through the polyimide sheet with 75 $\mu$m in thickness and 50 cm$^{2}$ in area during 60 min has been estimated as $\sim$1 $\times$ 10$^{-9}$ cm$^{3}$ STP, which corresponds to an ion counts of 1.3 $\times$ 10$^{5}$ cps if Ne is expanded to a volume of 1000 cc and a mass spectrometer with the sensitivity of 5 $\times$ 10$^{-3}$ (counts/sec)/(particle/cc) is used. For comparison, sensitivities of $\sim$5 $\times$ 10$^{-3}$ and $\sim$1 $\times$ 10$^{-2}$ (counts/sec)/(particle/cc) have been reported for the quadrupole mass spectrometer on the Curiosity rover \citep{Mahaffy+2012} and the Reflectron-type Time-Of-Flight mass spectrometer (RTOF) on the Rosetta Orbiter Spectrometer for Ion and Neutral Analysis (ROSINA) \citep{Balsiger+2007}, respectively. Sensitivities of 2 $\times$ 10$^{-3}$ for $^{4}$He and 3 $\times$ 10$^{-2}$ for $^{40}$Ar at the emission current of 250 $\mu$m and calibration procedures about Neutral Gas and Ion Mass Spectrometer (NGIMS) on MAVEN have been reported \citep{Mahaffy+2015}. The amount of $^{22}$Ne is likely $\sim$1/10 of $^{20}$Ne (Section \ref{subsec:knowledge}), so that the ion count of $^{22}$Ne is estimated to be $\sim$1 $\times$ 10$^{4}$ cps. The uncertainties for $^{20}$Ne/$^{22}$Ne ratios are arisen from a statistical error and corrections for mass discrimination, blank, and doubly charged ions ($^{40}$Ar$^{++}$ and CO$_2^{++}$). The statistical error for the isotope ratio is $\sim$1\% in the case of the above amount of $^{20}$Ne permeated. The contributions of blank $^{20}$Ne, $^{40}$Ar$^{++}$, and CO$_2^{++}$ are considered to be about 10\%, 10\%, and a few \% of permeating $^{20}$Ne (for $^{20}$Ne and $^{40}$Ar$^{++}$) or $^{22}$Ne (for CO$_2^{++}$), respectively, when the amounts of blank are assumed to be similar to those in the laboratory experiment performed by \citet{Miura+2020}. Because the source of the blank $^{20}$Ne and the contribution of $^{40}$Ar$^{++}$ are from the terrestrial atmosphere dissolved into a sealing material made from Viton, these values are likely upper limits and can be improved by revising the sealing material and/or considering the pressure of the Martian atmosphere being two orders of magnitude lower than that of the terrestrial atmosphere. Since CO$_{2}$ is mostly removed using a getter, the main contribution is of the background CO$_{2}$ from the ion source of the mass spectrometer. Assessing these error factors, where 50\% uncertainty to each correction is assumed, the $^{20}$Ne/$^{22}$Ne ratio is expected to be obtained with $<$10\% uncertainty. Therefore, this methodology for in situ mass spectrometry utilizing a permeable membrane can identify either solar-nebular- or implanted-solar-wind-like Ne or chondritic Ne as the source of Martian atmospheric Ne (Figure \ref{fig:isotopes}). 

Distinguishing captured-solar-nebula ($^{20}$Ne/$^{22}$Ne = 13.36$\pm$0.16) and implanted-solar-wind ($^{20}$Ne/$^{22}$Ne $\sim$ = 12.52--12.7) components is more challenging, and requires the determination of the atmospheric Ne isotopic ratio with $<$1\% uncertainty (Figure \ref{fig:isotopes}). This might be possible by adopting a mass spectrometer with high sensitivity and/or high resolution, developing a method for more efficient Ne-Ar separation, tuning of ion source parameters to lower production rates of the doubly charged ions \citep[e.g., a lower electron energy reduces ++/+ ratios;][]{Meshik+2012, Miura+2020}, utilizing a wider membrane area and/or better reduction of blank. It is also crucial to improve the model for Ne escape to reduce the uncertainty in isotopic fractionation factor, which directly influences the uncertainty in converting the atmospheric to mantle Ne isotopic ratios (Equation \ref{eq:20Ne22Ne}).

We utilized the absolute abundance of $^{20}$Ne in the Martian atmosphere estimated from $^{22}$Ne abundance measured with Viking (Figure \ref{fig:abundances}), but direct determination of $^{20}$Ne abundance is desired. For this purpose, in situ calibration of the mass spectrometer during the mission and accurate determination of the permeability and diffusion coefficient using a flight-model membrane in a laboratory are essential. Based on their experience at the laboratory, \citet{Blanchard+1986} reported that operation of the quadrupole mass spectrometer system with tungsten filaments at constant temperature has yielded more stable operation with weekly sensitivity changes being less than 10\%. Even for a space mission, it is recommended to carry a calibration gas for Ne to obtain a reliable $^{20}$Ne abundance. The calibration-gas-system consists of a gas reservoir and some micro valves as utilized in the Curiosity and MAVEN missions \citep{Mahaffy+2015}. In addition, the linearity of signals against gas amounts in the mass spectrometer must be examined carefully, although noble gases generally show a liner trend over several orders and the amount of calibration gas should be controlled to come to similar amounts between calibration gas and sample gas. As a result, we expect that the abundance of Ne will be determined within the uncertainty of a factor of two, which will improve the estimate for the mantle Ne abundance and constrain its source (Figure \ref{fig:abundances}).

Mars Sample Return \citep[MSR,][]{Beaty+2019} is another potential opportunity to measure the elemental and isotopic abundance of Ne on Mars. Rock samples potentially contain atmospheric and/or mantle-derived volatiles in trapped gas as we know it from Martian meteorites. However, both Martian meteorites and surface rocks analyzed in situ by Curiosity showed the dominance of cosmogenic Ne \citep{Farley+2014,Smith+2020}, which inhibited the determination of atmospheric or mantle-derived components. The possibility to recover Mars' atmospheric and/or mantle Ne endmembers would be dependent on the properties of returned rock samples such as the intrinsic to cosmogenic gas ratio and the total amount of trapped gas. Sample return of atmospheric gas itself would be, If performed, a more promising opportunity to measure Ne abundance in the Martian atmosphere.

\subsection{Constraints from other volatile elements}

$^{20}$Ne/$^{22}$Ne ratio may not solely provide conclusive determination of the origin of abundant mantle Ne especially if Ne was sourced from multiple materials. Thus, combining multiple volatile elements is useful for constrain the origin of abundant mantle Ne.

Isotopic compositions of H and nitrogen (N) in the Martian mantle are informed to be broadly chondritic \citep{Usui+2012,Mathew+Marti2001}. This would be naturally reconciled if the mantle Ne also originated from chondritic materials. The chondritic origin for H and N in the Martian mantle would limit the contribution of comets, whose D/H and $^{15}$N/$^{14}$N ratios are distinctly higher than chondrites \citep{Furi+Marty2015}, though the determination of bulk Ne abundance in comets is needed for further quantification. In contrast, solar nebula gas has D/H and $^{15}$N/$^{14}$N ratios lower than chondrites \citep{Furi+Marty2015}. This may limit the contribution of solar nebula gas on Martian volatiles, though mixing with cometary volatiles and/or isotopic fractionation via atmospheric escape \citep[e.g.,][]{Genda+Ikoma2008,Lammer+2020} can elevate these ratios.

\citet{Hutchins+1997} estimated $^{20}$Ne/$^{36}$Ar degassing ratio to be 10--26 by modeling the evolution of Ne and Ar isotopic ratios in the Martian atmosphere. If the ratio reflects the incorporation during the magma ocean stage, multiplying their solubilities \citep[1.6 ppm/MPa vs. 0.2 ppm/MPa from the model of][]{Iacono-Marziano+2010} leads to the original (source) ratio of $^{20}$Ne/$^{36}$Ar $\simeq$ 1.3--3.3. This ratio is lower than that of solar gas \citep[$\sim$10,][]{Lodders2003}, and thus suggests the contribution of chondritic, cometary, or IDP-like materials. Though the estimated degassing ratio may not be applicable if we consider Ar supply by late accretion \citep{Kurokawa+2018,Sakuraba+2019}, combining multiple noble gases is potentially useful for constraining the origin of Ne.

\subsection{Implications for the surface environment of early Mars}

We showed that a large amount of Ne in the atmosphere might exist on Mars when it formed. The lower limit of Ne partial pressure is $\simeq$10 Pa, which is higher than the current value (0.6--4 $\times$ 10$^{-3}$ Pa) by about four orders of magnitude. This excess amount of atmospheric Ne can be easily removed by atmospheric escape processes such as hydrodynamic escape \citep{Pepin1991}, impact erosion \citep{Melosh+Vickery1989,Sakuraba+2019}, and sputtering \citep{Jakosky+1994}. However, the source of rich Ne likely supplied major volatile elements (H, C, and N) to form a dense atmosphere. Depending on its amount, such a dense atmosphere might be left for $\sim$100 Myrs timescale. For instance, a recent study \citep{Yoshida+Kuramoto2020} suggested that the reducing remnants of Mars' primordial atmosphere, if existed, may contribute to warm early Mars \citep{Ramirez+2014,Ramirez2017,Wordsworth+2017} and to form organic matters found to be preserved at Gale crater \citep{Eigenbrode+2018}. In contrast, if the abundant Ne was sourced by chondritic or cometary matters, bulk Mars and, consequently, its early atmosphere could be more oxidizing. Secular volcanic degassing of these volatile elements once trapped in the solidified mantle would have also influenced the evolution of early Mars. Thus, unveiling the origin of abundant Ne in Mars' mantle is critical for understanding its early surface environment.

\subsection{Earth vs. Mars}
\label{subsec:EarthMars}

Comparing mantle Ne abundances and isotopic ratios between Earth and Mars (Section \ref{subsec:mantleNe}) provides several implications for terrestrial planet formation. Substituting Earth's mantle Ne abundance into Equation \ref{eq:PNerequired} results in the lower limit of required Ne partial pressure in the magma ocean stage to be $\simeq$2 Pa. In contrast to Mars, the estimate shows a good match with Ne partial pressure of the current atmosphere \citep[1.8 Pa,][]{Catling+Kasting2017}. Late processing such as hydrodynamic escape \citep{Hunten+1987,Pepin1991,Dauphas2003} and impact erosion and replenishment \citep{Marty+Meibom2007,Sakuraba+2019} likely influenced Earth's atmospheric Ne content. Moreover, the Earth's primitive mantle could possess more Ne than today's. For example, Earth's depleted mantle has $\simeq$ 1$\times$10$^{-13}$ g/g \citep{Moreira+Kurz2013,Marty2012}, and it has been proposed to be 100 times higher in the beginning \citep{Jaupart+2017}. Thus, the consistency of partitioning model for Earth might be a coincidence. However, it emphasizes that Mars' mantle Ne content requires an explanation.

$^{20}$Ne/$^{22}$Ne ratio in the Martian mantle is unknown. The ratio in Earth's deep mantle is close to the solar value and suggests the solar nebula origin \citep{Williams+Mukhopadhyay2019}, while the solar-wind-implanted dust has also been proposed as the origin of Ne \citep{Moreira+2016}. In the solar nebula origin scenario, Earth's accretion timescale longer than that of Mars \citep{Kleine+2009,Dauphas+Pourmand2011} as well as the larger mass can result in the different amount of nebula gas capture (Section \ref{subsec:capture}). In the case of the solar-wind-implanted dust, there is no reason to expect that Mars preferentially accreted this component compared to Earth. Multiple degassing events followed by atmospheric loss possibly due to giant impacts on Earth \citep{Tucker+Mukhopadhyay2014} may lead to the difference to Mars. 

\section{Conclusions}
\label{sec:conclusions}

Ne abundance in the Martian atmosphere was estimated from in situ measurements with Viking landers to be 1--6 ppm. Combining this abundance to the results of numerical modeling validated by the recent upper atmospheric measurements with MAVEN, we estimated the sputtering loss rate and the lifetime of Ne in Martian atmosphere to be 0.7--10 $\times$ 10$^{20}$ s$^{-1}$ and 0.6--1 $\times$ 10$^8$ years, respectively. The short lifetime implies recent or ongoing supply to the atmosphere, but accretion of asteroids, comets, and IDPs and cosmogenic production were found to insufficient to balance the loss. Thus, we proposed that the presence of atmospheric Ne is evidence of recent or ongoing volcanism on Mars. Ne abundance in the mantle, estimated by using photographic estimate of the magma production rate, is >0.5--8 $\times$ 10$^{-11}$ g/g, which is >5--80 times greater than that in current Earth's mantle. Furthermore, considering dissolution equilibrium between the molten mantle and the overlying atmosphere in the magma ocean stage showed that >300 times more Ne is required to put the estimated Ne abundance into the mantle. 

Several possible origins of Martian mantle Ne was discussed. We showed that the solar nebula gas capture under typical conditions is insufficient to explain the abundant Ne in the mantle. Mixing of a degassed component and completion of accretion prior to the nebula gas dissipation were proposed for efficient capture of the nebula gas. On the other hand, Ne can be supplied by chondrites, comets, and solar-wind-implanted dust. These possible Ne sources constrain Mars' accretion history and would have also supplied other highly volatile elements such as H, C, N, and S. Secular degassing from such a volatile-rich mantle has likely affected Mars' climate evolution. Finally, a simple relation between atmospheric and mantle $^{20}$Ne/$^{22}$Ne ratios were provided to constrain the origin of mantle Ne by measurements of atmospheric $^{20}$Ne/$^{22}$Ne ratio with a future Mars exploration mission or with Mars Sample Return.

\section*{Acknowledgments}

We thank Bruce Jakosky and an anonymous reviewer for constructive comments. This study was supported by JSPS KAKENHI Grant numbers 17H01175, 17H06457, 18K13602, 19H01960, 19H05072, JP20H0019, and 21K13976 and JSPS Core-to-Core program "International Network of Planetary Sciences."

\section*{Data availability}

All data used in this study are given in the main text. Analytic calculations to support the figures are all presented in the main text and an Appendix.

\section*{Author contributions}

All authors designed the project and wrote the manuscript. H.K. performed analytic calculations.

\appendix

\section{Analytical solution for the trapped Ne mass in the mantle} \label{ap:MNe}

Here we consider partitioning of Ne between the atmosphere, magma ocean, and solidified mantle, and derive an analytical solution for the Ne mass in the mantle at the time when bottom-up mantle solidification is completed. The mass-balance equations for Ne and mantle are given by,
\begin{equation}
    M_{\rm Ne, tot} = M_{\rm Ne, atm} + M_{\rm Ne, mo} + M_{\rm Ne, sm}, \label{eq:mass-balanceNe}
\end{equation}
and,
\begin{equation}
    M_{\rm mantle} = M_{\rm mo} + M_{\rm sm}. \label{eq:mass-balance}
\end{equation}
Combining Equations \ref{eq:dMNe}, \ref{eq:mass-balanceNe}, and \ref{eq:mass-balance}, we obtain, 
\begin{equation}
    \frac{dM_{\rm Ne, sm}}{M_{\rm Ne, tot}-M_{\rm Ne, sm}} = \frac{F_{\rm tl} dM_{\rm sm}}{m_{\rm Ne}A/\overline{m} g S_{\rm Ne} + M_{\rm mantle} - M_{\rm sm}}. \label{eq:dMdM}
\end{equation}
Integrating Equation \ref{eq:dMdM} leads to Equation \ref{eq:MNemantle}.

%\end{linenumbers}

%% References



%% New version of the num-names style
%\bibliographystyle{elsarticle-num-names} % original
\bibliographystyle{elsarticle-harv}\biboptions{authoryear} % HK's preference
\bibliography{references.bib}






% \end{thebibliography}

\end{document}